\documentstyle[aps,prl,graphicx]{revtex}

\begin{document}
\draft

\twocolumn[\hsize\textwidth\columnwidth\hsize\csname @twocolumnfalse\endcsname

\title{ Replica symmetry breaking in  long-range glass models without 
                   quenched disorder.}
\author{L.B. Ioffe, A.V. Lopatin}
\address{Department of Physics, Rutgers University, Piscataway, 
New Jersey  08855}
\date{\today}
\maketitle

\begin{abstract}
We discuss mean field theory of glasses without quenched disorder  focusing on 
the justification of the replica approach to thermodynamics. We emphasize the
assumptions implicit in this method and discuss how they can be verified. The 
formalism is applied to the long range Ising model with orthogonal coupling matrix.
 We find the 
one step replica-symmetry breaking solution and show that it is stable in the 
intermediate temperature range that includes the glass state but excludes 
very low temperatures. At very low temperatures this solution becomes unstable 
and this approach fails.

\end{abstract}

\vskip2pc]

\bigskip

The thermodynamics of glasses without quenched disorder is a long standing
problem in statistical physics. The interest to this problem was renewed
recently when it was understood that powerful methods developed for the
glasses with quenched disorder can be often applied to this problem. 
\cite{Marinari94,Chandra95,monasson,franz,mezard1,Mezard2}. In both systems the
local magnetization in the ground state varies from site to site and different 
sites are typically non-equivalent. The qualitative reason why glasses without 
quenched disorder are more difficult to describe theoretically than spin glasses 
is the following. The mean field theory has to operate with the average 
magnetization (or its moments), not with quantities which depend on a 
realization and a particular state. The average quantities appear naturally 
in spin glasses after averaging over quenched disorder which makes all 
sites equivalent.

A few methods were suggested to overcome this difficulty for the glasses
without quenched disorder. First, a mapping of some glass models to the
quenched disordered problems was suggested \cite{Marinari94}, this method has an
obvious disadvantage that such mapping is difficult to guess. Second, it was
noted that a typical dynamics in a glassy system leads not to a ground state
but to one of many metastable states providing an effective averaging
mechanism \cite{Chandra96} which makes all sites equivalent even for glasses
without quenched disorder. This method has a disadvantage that dynamical
equations are much more difficult to solve than the statical ones. Very
recently the cloning method was proposed that is based on the idea that even
at low $T$ a system of $m$ clones might be distributed in its phase space
over many low lying metastable states if $m$ is chosen correctly and the
properties of all these states are essentially equivalent to those of the
ground state. \cite{monasson,franz,mezard1,Mezard2}
Generally, the partition sum of $m$ weakly coupled clones is $%
\sum_{F}e^{-N(m\beta F-S_{conf}(F))}$, where sum goes over free energies
(per site)\ of metastable states, $F$, and $S_{conf}(F)$ is their
configurational entropy ($S_{conf}=\frac{1}{N}\ln ({\cal N}_{states})$).
Assuming that $dS_{conf}(F)/dF$ is finite at the lowest $F$ associated with
the ground state one needs to chose $m\propto T$ at low $T$ in order to
avoid a complete dominance by a single (ground) state and the problems with
site non-equivalence mentioned above. Distributing the system in the phase
space provides the effective averaging mechanism in this approach. The main
assumptions implicit in this approach are that low lying metastable states
are not correlated (otherwise, averaging over them would not remove
completely the non-equivalence of different sites) and that configurational
entropy associated with these states behaves well as a function of energy at
low energies permitting the ''right'' choice of $m$.

The goal of this paper is to provide an alternative theoretical framework to
the cloning method which, albeit somewhat similar in formalism, uses
different physical arguments for its justification and allows to check the
main assumption of the method mentioned above. We apply this method
to the Ising version of the periodic long-range Josephson array model which
is a simple example of a glass without quenched disorder and we show that in
this model the main assumptions of the method are correct in the
intermediate temperature range but become wrong at very low temperatures.
The main idea of the approach is that in a system with many low-lying states
even a small random field is able to change the energy balance between the
states and pull down a different state making it a new ground state of the
system. Averaging over this random field is equivalent to the averaging over
low lying metastable states. Specifically, in a spin system we add to the
physical Hamiltonian a magnetic field part: ${\cal H}\rightarrow {\cal H}%
+\sum_{i}h_{i}S_{i}$ with small random $h_{i}$. The resulting change in the
energy of a typical metastable state is of the order of$\sqrt{N}h$; because
this energy interval contains a large amount of metastable states, we expect
that a small non-zero field would result in a large rearrangement of their
energies but would not change the properties of individual states. Averaging
over the random field configurations is performed in the usual way
introducing $n$ replicas of the system and taking the limit $n\rightarrow 0$. 
The assumption of uncorrelated states is equivalent to one step replica
symmetry breaking (1RSB) formalism; further, in this case this method is
formally equivalent to cloned liquid approach if the size of the blocks in
1RSB is equal to the number of clones (see \cite{Mezard2} for the discussion
of replica method vs clones for quenched disordered glasses). From the above
discussion it is evident that another assumption implicit in this approach
is that the energy spacing between low lying states should be much less
than $O(\sqrt{N})$ if it is too big
a small magnetic field will not be sufficient to rearrange low lying
states, if it is too small, e.g. $d S_{conf}/dF|_{F_0}=\infty,$ the effect 
of random field will be too large and no sensible limit $h_{i}\rightarrow 0$ 
is possible. The latter situation seems to happen in the periodic long range 
Josephson array with flux $2\pi $ per strip \cite{Chandra95} when all states 
are exactly degenerate and $S_{conf}(F)$ is very singular at $T=0$.

We now provide the details of our formalism and its application to the
simplest model of a glass without disorder. Our model consists of two sets
of Ising spins (which we shall refer as ''upper'' and ''lower'' in the
following) interacting via 
\begin{equation}
{\cal H}=-{\frac{1}{2}}\sum_{m,n}S_{im}J_{mn}^{ij}S_{jn},  \label{H}
\end{equation}
Here the spin $S_{im}$ has a site index ($m=1\ldots N$) and a components
index $i=1,2$ corresponding to the upper and lower spins and matrix $\hat{J} 
$ is

\begin{equation}
\hat{J}_{mn}=\left( 
\begin{array}{cc}
0 & J_{mn} \\ 
J_{mn} & 0
\end{array}
\right) ,  \label{J}
\end{equation}
with $J_{mn}=(J_{0}\sqrt{2/N})\cos (\frac{2\pi \alpha }{N}(m-1/2)(n-1/2))$.
For $\alpha =1/2$ we obtain the orthogonal limit $%
\sum_{n}J_{mn}J_{nk}=J_{0}^{2}\delta _{mk}$, in what follows we shall focus
on this case. This Ising spin model is similar to the $XY$ spin model of
long ranged Josephson array \cite{Chandra95} and to the
Bernasconi model \cite{Bernasconi}. As well as in these models
its lowest states correspond to ``pseudorandom'' sequences
with flat Fourier transform.  So, we expect that this model also
displays glassy properties, in particular that it has extensive
configurational entropy at low temperatures. Further, one expects that in a
model with long range interaction the barriers separating metastable states
become infinite in the thermodynamic limit. We have verified numerically
that the configurational entropy in this model is indeed extensive and its
dependence on energy is similar to the one obtained for other infinite range
glasses (see Fig. 1). Note, however, the important
difference between this model and the $XY$ spin model of \cite{Chandra95}:
in the orthogonal limit the ground state of Ising model does not become
extensively degenerate (i.e. degeneracy stays finite as $N \rightarrow
\infty $, see Fig. 1) whereas in the $XY$ spin model the ground state
becomes extensively degenerate in the unitary limit making it very
complicated \cite{Chandra95}.

Taking the Gaussian distribution for the random magnetic field $\langle
h_{i}\,h_{j}\rangle =2\,h_{0}^{2}\,\delta _{i,j}$ we get the replica
Hamiltonian 
\begin{equation}
{\cal H}_{s}=\sum_{\alpha }{\cal H}(S_{\alpha })+h_{0}^{2}\sum_{\alpha
,\beta .i}S_{im}^{\alpha }S_{im}^{\beta },  \label{Hs}
\end{equation}
where the replica indexes $\alpha ,\beta $ run from $1$ to $n$ and the limit 
$n\rightarrow 0$ should be taken. The glass transition corresponds to an
appearance of a non replica-symmetric solution of the Hamiltonian (\ref{Hs})
in the limit $h_{0}\rightarrow 0.$

\begin{figure}[ht]
\includegraphics[width=3.2in]{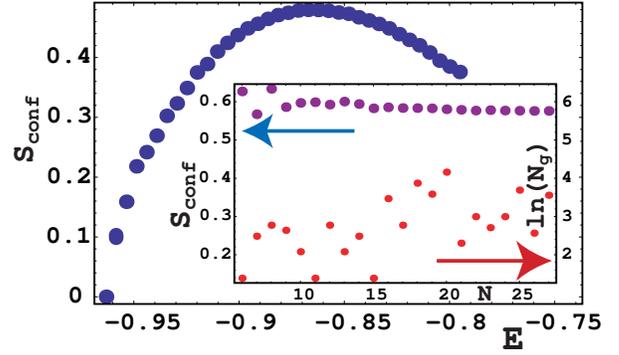}
\caption{Main panel: configurational entropy $S_{conf}=\ln({\cal N})/N$
at $T=0$ as a function of state energy obtained from direct numerics on systems
up to $N=27$ size.  Insert:  total $S_{conf}$ and the degeneracy, $N_g$, of the lowest
energy state at $T=0$ as a function of system size.}
\end{figure}

In the large $N$ limit a long range model containing $N$ sites can
be reduced to an effective single-site model with a free energy density $%
{\cal F}$ 
\begin{equation}
-\beta {\cal F}={\frac{1}{2}}{\rm Tr}\,\gamma (B)+{1\over 2}\sum_{j}S_{j}^{\alpha
}B_{\alpha \beta }S_{j}^{\beta },  \label{F}
\end{equation}
where $S_{j}^{\alpha }$ is Ising spin field retaining only replica and
component index dependence, $B$ is an order parameter matrix in the replica
space. The function $\gamma (B)$ can be determined from the condition that
all single site correlation functions of the model (\ref{F}) coincide with
the correlation functions of the original model (\ref{Hs}). Instead of
comparing the spin correlation functions of these two models it is easier to
decouple Ising spins by auxiliary field $\psi $, sum over Ising spins and
compare the correlation functions of conjugate field $\psi $ in the two new
models

\begin{equation}
\beta {{\cal H}_{\psi }}={\frac{T}{2}}\sum_{m,n,\alpha }\psi _{im}^{\alpha }(%
\hat{J}^{-1})_{mn}^{ij}\psi _{jn}^{\alpha }-\sum_{m,\alpha ,j} V(\psi
_{jm}^{\alpha }),  \label{Hpsi}
\end{equation}

\begin{eqnarray}
\beta {\cal F}_{\psi } = -\frac{1}{2} \left[ {\rm Tr}\,\gamma (B) \! -
\sum_{\alpha ,\beta ,j}\psi _{j}^{\alpha }B_{\alpha \beta }^{-1}\psi
_{j}^{\beta } \right] \!-\! \sum_{\alpha ,j}V(\psi _{j}^{\alpha }).
\label{Fpsi}
\end{eqnarray}
where $V(\psi)=\ln 2\cosh (\psi)$. For both models one can construct a
formal perturbation theory in the interaction $\ln 2\cosh (\psi _{\alpha
}^{(j)})$ and verify that these expansions coincide. 
We begin with the model (\ref{Hpsi}). Inspecting the terms of the
perturbation theory for the correlator $G_{im,jn}^{\alpha \beta }=\langle
\psi _{im}^{\alpha }\psi _{jn}^{\beta }\rangle $ one verifies
that in the leading order in $1/N$ it is given by $\hat{G}=[T\hat{J}%
^{-1}-\Sigma ]^{-1}$ with the self energy $\Sigma $ which is diagonal in the
site index: $\Sigma ={\cal A}\,\delta _{mn}\,\delta_{i,j}$. This approach is similar 
to a locator expansion \cite{Feigelman} but in our case the locator ${\cal A}$ 
might be non-trivial in the replica space.  Using the orthogonality of $\hat{J}$ we 
get that the single site correlation function ${\cal G}_{\alpha \beta }\equiv
G_{im,im}^{\alpha \beta }$ (that we need to establish the correspondence
between the models) becomes 
\begin{equation}
{\cal G}=[-{\cal A}+(j_{0}^{2}{\cal A})^{-1}]^{-1}  \label{G}
\end{equation}
where $j_{0}=\beta J$.

Now we turn to the model (\ref{Fpsi}). Here the self energy is diagonal in
the site index by construction, further, the interaction part of this model
is the same as for model (\ref{Hpsi}); assuming that their single site
correlation functions coincide we conclude that their single site
self-energies are equal as well. Thus, the spin correlator obtained for this
model is ${\cal G}=[B^{-1}-{\cal A}]^{-1}$, comparing this expression with (%
\ref{G}) we conclude that $B=j_{0}^{2}{\cal A}$ .

The correlator of the dual field $\psi $ can be related to the correlator of
original spins: consider a Gaussian transformation leading to the model (\ref
{Fpsi}) $\exp (SBS/2)=\int d\psi \exp (-\psi B^{-1}\psi/2 +S\psi )$ and use it
to express ${\cal G}$ via correlator $D_{\alpha \beta }\equiv \langle
S^{\alpha }S^{\beta }\rangle $, we get: ${\cal G=}B+BDB$. Solving this
equation for the spin correlator $D$ and using (\ref{G}) and the relation $%
B=j_{0}^{2}{\cal A}$ we obtain 
\begin{equation}
D=B[j_{0}^{2}-B^{2}]^{-1}.  \label{D}
\end{equation}
Finally, the saddle point condition for the free energy (\ref{F}) $%
2D=-\gamma ^{\prime }(B),$ therefore integrating Eq.(\ref{D}) we find

\begin{equation}
\gamma (B)=\ln (1-j_{0}^{-2}B^{2}),  \label{gamma}
\end{equation}
Note that the free energy (\ref{F}) coincides with the free energy of the 
model considered in Ref.\cite{Marinari94} although their properties at finite $N$ 
are markedly different. Furthermore, this free energy is the same as obtained by 
fiduciary Hamiltonian approach \cite{Marinari94}.

{\it Paramagnetic state.} In this state we take the replica symmetric ansatz 
$B_{\alpha ,\beta }=\mu \,\delta _{\alpha ,\beta }$ and free energy (\ref{F}%
) becomes 
\begin{equation}
{\cal F}/T=[\ln j_{0}^{2}-\ln (j_{0}^{2}-\mu ^{2})]/2-\mu -2\ln 2.
\end{equation}
Variation with respect to $\mu $ gives $\mu =[\sqrt{1+4j_{0}^{2}}-1]/2.$
Usual thermodynamic relations between energy and entropy give 
$E=-T\,\mu, \;S=\ln [4\sqrt{\mu }/j_{0}].$

One can see that the entropy of the normal solution becomes negative at $%
T<T_{K}=J_{0}/(4\sqrt{15})\approx 0.064550\;J_{0}$ which is the Kauzmann
temperature for this model \cite{Kauzmann48}; one expects that the glass
transition takes place at some temperature, $T_c$, above $T_{K}$.

{\it Glass state.} At the glass transition temperature $T_{c}$ the replica symmetry
is broken, we assume that it is described by one step replica symmetry breaking (1RSB)
and then verify that it is indeed a stable solution below $T_c$. 
%
The 1RSB ansatz is $B_{\alpha ,\beta }=\mu \,\delta _{\alpha ,\beta }+\eta
\,R_{\alpha ,\beta },$ where the matrix $R$ is a block-diagonal matrix
consisting of $m \times m$ blocks with all elements equal $1$, we get the
free energy functional 
\[
\hspace{-1.5cm}\beta{\cal F}=[\log \,j_{0}^{2}-(1-1/m)\,\ln \,(j_{0}^{2}-\mu
^{2})]/2
\]
\begin{equation}
-2\ln 2-(\ln X)/2m-\mu -2\,f(\eta ,m)/m,
\end{equation}
where $X=j_{0}^{2}-(\mu +\eta m)^{2}$ and the function $f$ is 
\begin{equation}
f(\eta ,m)=\ln \left[ \int P_m(z) dz\right], \; P_m(z)= \frac{e^{-z^{2}/2}}{\sqrt{2\pi }}
\cosh^{m}(z\sqrt{\eta }).
\end{equation}
Taking the derivatives of ${\cal F}$ with respect to $\mu ,\eta ,m$ we get 
\begin{equation}
\left( {\frac{1}{m}}-1\right) {\frac{\mu }{{j_{0}^{2}-\mu ^{2}}}}-{\frac{{%
\eta m+\mu }}{Xm}}+1=0,  \label{first}
\end{equation}
\begin{equation}
-(\eta m+\mu )/X+q(m-1)+1=0,  \label{second}
\end{equation}
\[
\hspace{-1cm}{\frac{1}{{2m^{2}}}}\log \left[ (j_{0}^{2}-\mu ^{2})/X\right] +{%
\frac{2}{m}}\,{\frac{\partial }{{\partial m}}}f(\eta ,m)
\]
\begin{equation}
\hspace{1cm}-\eta \,(\mu + \eta m)/mX-2\,f(\eta ,m)/m^{2}=0,  \label{third}
\end{equation}
where $q=\int \tanh^2(z) P_m(z)dz /\int P_m(z)dz$  
is the spin overlap of different replicas belonging to the same
block $D_{\alpha ,\beta }=(1-q)\delta _{\alpha ,\beta }+q\,R_{\alpha ,\beta
},$ that coincides with  Edwards-Anderson (EA) order parameter. Eqs.(\ref
{first},\ref{second}) can be solved with respect to $m,\eta $ giving 
\begin{equation}
\mu  =\eta {\frac{{1+(1-q+qm)\eta m}}{{q/(1-q)-2\eta \,(1-q+qm)}}}
\end{equation}
and $j_{0}^{2} =\mu ^{2}+\mu /(1-q).$
For a given $m$ we can solve Eq.(\ref{third}) numerically with respect to $%
\eta $ and get all quantities as functions of $m.$ The resulting dependence
of  $m(T)$ for $J_{0}=1$ is shown on Fig.~2. In the limit $n\rightarrow 0,$
the values of $m$ should lie within the interval $(0,1)$ and $m=1$ defines
the thermodynamic critical temperature $T_{c}\approx 0.064593$, it is larger
than $T_{K}$ as expected. The value of the EA order parameter $q$ at the
glass transition is very close to $1,$ $1-q=0.00017116$, so in this sense
the phase transition is strongly first order but (similar to p-spin model),
the energy and entropy do not change discontinuously at the transition. The
numerical solution shows that when the temperature decreases, the entropy of
the glass state monotonically decreases and eventually becomes negative
below $T_K^\prime\approx 2.8\times 10^{-4}.$ The explanation of such unphysical
behavior is that 1RSB ansatz, in fact, becomes unstable in this low
temperature regime.

{\it Stability of the thermodynamical solution.} In order to analyze
stability of 1RSB ansatz we expand the Eq.(\ref{F}) to the second order in
fluctuation of the order parameter $\delta B$ and consider different
families of fluctuation matrices $\delta B$. This calculation is very similar to
the analysis of the stability of paramagnetic solution and Parisi solution in SK 
model \cite{Almeida,DeDominicis} so we only sketch it here, for details see Appendix. 
We find that the most dangerous direction in the fluctuation space corresponds
to the "replicon" modes \cite{Almeida,DeDominicis} that are fluctuations within 
diagonal blocks of $\delta B$  satisfying the conditions $(\delta  B\,
R)_{\alpha ,\beta }=0$, $\delta B_{\alpha ,\alpha }=0.$  The eigenvalue
corresponding to these modes is 
$$
\Lambda=2(1-q)/\mu+2(1-q)^{2}-2(r-q^{2})
$$
where 
$r = \int \tanh^4(z) P_m(z)dz /\int P_m(z) dz$. 
Numerical solution shows that $\Lambda _{1}$ is positive at
temperatures $T>T_{uns}\approx 6.1\times 10^{-3}$ but changes sign at $T_{uns}$, 
thus 1RSB solution is unstable at $T<T_{uns}.$ 

\begin{figure}[ht]
\includegraphics[width=3.2in]{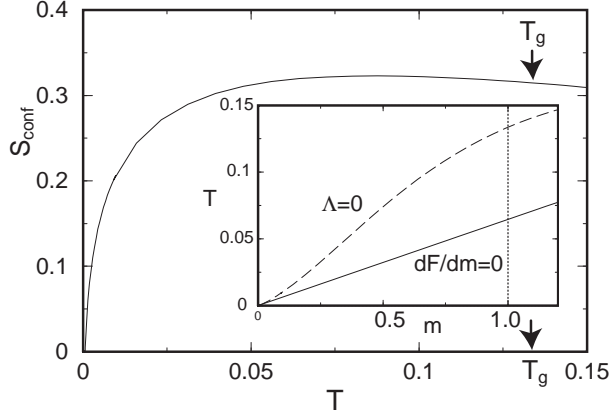}
\caption{Main plot: Dependence of the configuration entropy on the temperature for the
marginally stable solution. Insert: size of 1RSB block, $m$, for thermodynamic
(solid line) and marginally stable (dashed line) solutions. The value of $T$ at 
at which $m=1$ gives thermodynamical (dynamical) critical
temperature.}
\end{figure}

{\it Marginal solution.} One expects that in a glass a typical dynamical process leads to
a most abundant state which is, therefore, marginally stable. We note that
although plausible, this assumption might be violated if the attraction basins of the low
lying states are much larger than those of the marginally stable ones \cite{Chandra00}.
Assuming that it is not the case, a dynamical freezing leads to the states with  
$\Lambda=0$ instead of the states with the minimal free energy characterized by 
$\partial {\cal F}/\partial m=0$.    Thus, to get the properties of the states selected
in a "dynamical'' process we replace Eq.(\ref{third}) by $\Lambda=0$.
 The resulting dependence of the temperature on the size of 1RSB
block $m$ is shown in Fig.2. The value $m=1$ defines the ''dynamical''
critical temperature $T_{g}\approx 0.13363.$ The free energy functional 
(\ref{F}) corresponding to 1RSB ansatz is equivalent to the free energy
functional that is obtained in the cloned liquid approach with $m$
being equal to the number of the clones \cite{Mezard2}. 
The stability of the 1RSB solution
indicates that the main assumptions of this approach are correct  in some temperature
range below $T_g$ and, therefore, in this temperature range the 
configurational entropy is given by 
$
S_{conf}=m^{2}{\frac{{\partial {\cal F}}}{{\partial m}}}
$.
The dependence $S_{conf}$  on temperature is shown in Fig 2.
Decreasing the temperature, it first increases, goes through the maximum
at $T_m$, and eventually becomes negative at the temperature  $T_{uns},$
at which the thermodynamical solution becomes unstable.  It is not clear however 
that the 1RSB solution 
is a correct solution in the whole temperature range $T_{uns}<T<T_g$, 
on the contrary, it is quite likely that another solution is preferred by the system 
below some $T_c'<T_g$.  We have only indirect arguments for this: first, it 
seems unphysical that  $S_{conf}$ decreases with temperature decrease, usually
at lower temperatures additional states appear. Second, the obtained $S_{conf}$ does not
match the results of numerics if one believes that this solution remains correct at
$T<T_m$. Finally, note the analogy with higher temperatures: paramagnetic solution
is always stable but is eventually replaced by 1RSB solution. 

In conclusion we have justified the application of replica method to some systems
without quenched disorder and discussed situations in which cases it fails. We
identify two dangers: correlations between metastable states close to the
ground state and too large degeneracy of the ground state. We apply the
formalism to the periodic Ising spin model with orthogonal coupling matrix
and  find that it gives the same free energy as fiduciary Hamiltonian approach 
\cite{Marinari94}.
Further we show that it  works in the intermediate temperature
range but fails at low temperatures when metastable states become
correlated. Two questions remain open: whether generalization of this method
to continuous symmetry breaking would allow one to study the models with
correlated metastable states and what to do if the ground state of the model
is highly degenerate as it is, e.g., in the case of unitary coupling matrix.

\onecolumn

\section{Appendix. Stability analysis.}

Here we provide the details of the stability analysis of 1RSB solution.  We
start by expanding the Eq.(\ref{F}) to the second order in fluctuation of
the order parameter $\delta B$:

\[
\delta F={\frac{1}{4}}\,\delta B_{\alpha ,\beta }\,\,M_{\alpha ,\beta
,\gamma \,\delta }\,\,\delta B_{\gamma ,\delta },
\]
\begin{equation}
M_{\alpha ,\beta ,\gamma ,\delta }=a\,R_{\beta ,\gamma }\,R_{\delta
,\alpha }+b\,[\delta _{\beta ,\gamma }\,R_{\delta ,\alpha }+\delta _{\delta
,\alpha }\,R_{\beta ,\gamma }] +c\;\delta _{\alpha ,\gamma }\,\delta _{\delta ,\alpha } + D_{\alpha ,\beta
}D_{\gamma ,\delta } - \Gamma _{\alpha ,\beta ,\gamma ,\delta },
\end{equation}
where the coefficients $a,b,c$ come from the expansion of ${\rm Tr}\;\gamma
(B)/2$ 
\begin{equation}
a=2q^{2}-2[Y\eta /\mu +q]\,[Y\eta /(\mu +\eta m)-q],
\end{equation}
\begin{equation}
b=q[\mu ^{-1}+4(1-q)] + Y\eta \,[(1-q)\eta m-1]/[\mu ^{2}+\eta \mu m],
\end{equation}
\begin{equation}
\hspace{-2cm}c=2(1-q)\,[2(1-q)+\mu ^{-1}],
\end{equation}
with $Y=1-q+qm.$ The function $\Gamma $ is the four spin replica correlation function 
$\Gamma _{\alpha ,\beta ,\gamma ,\delta }=\langle S_{\alpha }S_{\beta
}S_{\gamma }S_{\delta }\rangle ,$ defined with respect to the Hamiltonian (%
\ref{F}). When at least two indexes, e.g. $\alpha$ and $\beta,$ are equal 
this function  is
\begin{equation}
\Gamma_{\alpha,\alpha,\gamma,\delta}=(1-q)\,\delta_{\gamma,\delta}+
q\,R_{\gamma,\delta}.
\end{equation} 
If $\alpha,\beta,\gamma,\delta$ are all different, the function $\Gamma$ is not zero 
only if either $\alpha,\beta,\gamma,\delta$ belong to the same block, or two of 
indexes (e.g. $\alpha,\beta$) belong to one block and the other two ($\gamma,\delta$) 
to another one. In the first case 
$
\Gamma_{\alpha,\beta,\gamma,\delta}=r,
$
where $r$ is
\begin{equation}
r=\int P_m(z)\tanh^4(z) dz/\int P_m(z) dz.
\end{equation}
In the second case, since $\alpha,\beta$ and $\gamma,\delta$ belong to 
different blocks 
$
\Gamma_{\alpha,\beta,\gamma,\delta}=q^2.
$

 The eigenvalue equation for the matrix $M$ is 
\begin{equation}
M_{\alpha ,\beta ,\gamma ,\delta }\;\delta B_{\gamma ,\delta }=\Lambda
\;\delta B_{\alpha ,\beta }.
\end{equation}
The eigenvalue equation can be simplified using that $\delta B$ is a symmetric matrix.
The resulting equation is   
\begin{eqnarray}
[b-2q(1-q)]\,(\delta B\, R+R\,\delta B)_{\alpha,\beta}+ (a-2q^2)
\,(R\,\delta B\, R)_{\alpha,\beta}
-2(4q-3r-1)\,\delta_{\alpha,\beta}\,\,\delta B_{\alpha,\alpha} \nonumber \\
\vspace{0.1cm}
+(3q^2-r)R_{\alpha,\gamma}\,\delta\,B_{\gamma,\delta}
R_{\gamma,\delta}R_{\delta,\beta}+
2(q-r)\,\Bigl[2\,\delta_{\alpha,\beta}(R\,\delta B)_{\alpha,\alpha}+
R_{\alpha,\beta}\delta B_{\alpha,\alpha}
+R_{\beta,\alpha} \delta B_{\beta,\beta}\Bigr]  \nonumber \\
+(q^2-r) \Bigl[\delta_{\alpha,\beta} R_{\beta,\gamma} \delta B_{\gamma,\gamma}
+2R_{\alpha,\beta}
\delta B_{\alpha,\beta}-2 R_{\alpha,\beta}(\delta B\, R)_{\alpha,\alpha}
-2 R_{\alpha,\beta}(\delta B\, R)_{\beta,\beta} \nonumber \\
-\delta_{\alpha,\beta}(R\,\delta B\, R)_{\alpha,\alpha}-R_{\alpha,\gamma}
\delta B_{\gamma,\gamma}
R_{\gamma,\beta}\Bigr]=[\Lambda+2(1-q)^2-c]\, \delta B_{\alpha,\beta}.
\end{eqnarray}
This equation has a block-diagonal structure, therefore one can divide
the fluctuation matrix $\delta B$ into blocks $\delta {\cal B}$ of $m\times m$
size and consider the fluctuations within each block independently.
Moreover, all eigenvalue equations corresponding to diagonal (off-diagonal)
blocks are equivalent. Therefore we are left with two cases: (I)
fluctuations within an off-diagonal block and (II) fluctuations within a
diagonal block.

 We begin our analysis of eigenvalues with the case (I) for which the  eigenvalue 
equation is reduced to 
\begin{equation}
(b-2q(1-q))(\delta B\,{\cal E}+R\,\delta B)_{\alpha,\beta}+(a-2q^2)
({\cal E}\,\delta B\,{\cal E})_{\alpha,\beta}=
\Lambda^\prime\delta B_{\alpha,\beta}, \label{offdiag}
\end{equation}
 where $\Lambda^\prime=\Lambda-c+2(1-q)^2$ and ${\cal E}$ is the $m\times m$ 
matrix with all
elements equal 1.
 The first eigenvalue corresponds to $\delta B$ satisfying 
$(\delta {\cal B}\,{\cal E})_{\alpha,\beta}=0$ and it is
\begin{equation}
\Lambda_1^{(1)}=c-2(1-q)^2.
\end{equation}
 The second eigenvalue is 
\begin{equation}
\Lambda_2^{(1)}=c+mb-2Y(1-q)
\end{equation}
and it corresponds to  $\delta {\cal B}$ satisfying  
$
(\delta 
{\cal B}\,{\cal E})_{\alpha,\beta}\ne 0, \,\,({\cal E} \delta {\cal B}\,%
{\cal E})_{\alpha,\beta}=0.
$ 
 The last eigenvalue of type I  corresponds to $\delta {\cal B}
={\cal E}$
and it is 
\begin{equation}
\Lambda_3^{(1)}=c+mG-2Y^2,
\end{equation}
where $G=ma+2b.$

 The eigenvalues of type II satisfy the equation
\begin{eqnarray}
{\cal E}_{\alpha,\beta}\Bigl[2(q-r)(\delta{\cal B}_{\alpha,\alpha}
+\delta {\cal B}_{\beta,\beta})+2(r-q^2)[(\delta{\cal B}\,{\cal E})_{\alpha,\alpha}
+(\delta B\,{\cal E})_{\beta,\beta}]+(r-q^2){\rm Tr}\delta{\cal B} \Bigr]
\nonumber \\
-\delta_{\alpha,\beta}\Bigl[2(4q-3r-1)\delta{\cal B}_{\alpha,\alpha}
+4(r-q)({\cal E}\,\delta{\cal B})_{\alpha,\alpha}
+(r-q^2)[{\rm Tr}\delta{\cal B}
-({\cal E}\,\delta{\cal B}\,{\cal E})_{\alpha,\alpha}]\Bigr]
 \nonumber \\
(-2q (1-q)+b)(\delta {\cal B}\,{\cal E}+{\cal E}\,\delta{\cal B})_{\alpha,\beta}
+(a-r+q^2)({\cal E}\,\delta {\cal B}\,{\cal E})_{\alpha,\beta} 
=[\Lambda^\prime+2(r-q^2)]\delta {\cal B}_{\alpha,\beta}
\end{eqnarray}
 The first eigenvalue of this type corresponds to 
$(\delta {\cal B}\,{\cal E})_{\alpha,\beta}=0$
and it is 
\begin{equation}
\Lambda_1^{(2)}=c-2(1-q)^2-2(r-q^2).
\end{equation}
 All other eigenvalues of type II correspond to $\delta{\cal
B}$ of the form $\delta{\cal B}_{\alpha,\beta}=x_{\alpha}+y_{\beta}\,{\cal E}%
_{\alpha,\beta} +y_{\alpha}\, {\cal E}_{\alpha,\beta}, $ where $%
x_\alpha,\,y_\alpha$ are such that either $\sum_\alpha x_\alpha= \sum_\alpha
y_\alpha=0$ or $x_\alpha=x, y_\alpha=y.$ The corresponding eigenvalues are: 
$$
\Lambda_{2,3}^{(2)}=c+[mb+2E\pm\sqrt{(bm+2E)^2-16Eb}]/2, 
$$
$$
\Lambda_{4,5}^{(2)}=c+[mG+F\pm\sqrt{(mG+F)^2-4GF}]/2, 
$$
where $E=4q-3r-1+m(r-q)$ and $F=4[\partial^2/\partial
\eta^2]f(\eta,m)/(1-m)m.$
The numerical solution shows that for the thermodynamical solution all eigenvalues 
are positive for temperature higher than $T_{uns}.$ At temperature less than 
$T_{uns}$ the eigenvalue $\Lambda_1^{(2)}$ corresponding to the ``replicon''
mode becomes  negative. The marginal (``dynamical'') solution
is defined by $\Lambda_1^{(2)}=0,$ all the other eigenvalues for this solution
are strictly positive in the region where it exists. Thus the replicon mode
is always the most relevant fluctuation.

\end{document}